\documentclass[aps,prb,floats,twocolumn,superscriptaddress,floatfix]{revtex4}
\usepackage{epsf,graphicx}
\usepackage{amssymb}
\usepackage{amsmath}
\usepackage{eepic}
\usepackage{epstopdf}
\usepackage{color}
\usepackage{ulem}

\begin{document}

\title{Functional field integral approach to quantum work}
\author{Jian-Jun Dong}
\affiliation{Beijing National Laboratory for Condensed Matter Physics and Institute of
Physics, Chinese Academy of Sciences, Beijing 100190, China}
\affiliation{University of Chinese Academy of Sciences, Beijing 100049, China}
\author{Yi-feng Yang}
\email{yifeng@iphy.ac.cn}
\affiliation{Beijing National Laboratory for Condensed Matter Physics and Institute of
Physics, Chinese Academy of Sciences, Beijing 100190, China}
\affiliation{University of Chinese Academy of Sciences, Beijing 100049, China}
\affiliation{Songshan Lake Materials Laboratory, Dongguan, Guangdong 523808, China}
\affiliation{Collaborative Innovation Center of Quantum Matter, Beijing 100190, China}
\date{\today}

\begin{abstract}
We introduce the functional field integral approach to study the statistics of quantum work under nonequilibrium conditions and derive the general formalism for a bilinear Hamiltonian with arbitrary time dependence. The method is then examined in three models. For the transverse Ising chain, it yields the correct quantum critical scaling and dynamical quantum phase transitions for single and double quench protocols, respectively. For the Su-Schrieffer-Heeger (SSH) model, we observe nonuniversal quantum critical scaling with anomalous $1/N$-correction due to its topological nature. Dynamical quantum phase transitions are observed for three different time evolution protocols but their time periodicity only appears in the double quench case. We then extend our method to the Bardeen-Cooper-Schrieffer (BCS) model for superconductivity and discuss the possibility of its application for general correlated models in combination with either the mean-field approximation or exact Monte Carlo simulations on classical (auxiliary) fields or disorders. Our method has the advantage of numerical simplicity, in the cost of explicit state evolution, and provides a promising way for exploring the physics of quantum work under general conditions.
\end{abstract}
\maketitle

\section{Introduction}

Nonequilibrium conditions provide an additional dimension in time domain for probing the many-body dynamics beyond the well-established equilibrium statistics and have recently led to the proposal of many novel phenomena such as the dynamical quantum phase transition \cite{Heyl2013,Budich2016,Heyl2018,Lang2018}, the time crystal \cite{Wilczek2012,Else2016}, the fluctuation relations \cite{Talkner2007PRE,Talkner2007JPA,Crooks1999,Tasaki2000,Jarzynski1997PRL,Jarzynski1997PRE,Esposito2009,Campisi2011}, and so on. While ultracold atoms in optical lattices can be easily tuned to be out of equilibrium \cite{Bloch2008,Aoki2014}, recent development of ultrafast pump-probe spectroscopy has enabled one to study the excitation and relaxation dynamics in real correlated materials \cite{Wall2011}. The study of nonequilibrium quantum physics is becoming one of the most active and exciting branches of modern condensed matter physics and attracted intensive attentions in recent years \cite{Polkovnikov2011}. However, despite of many
theoretical progresses including the nonequilibrium extension of the density matrix renormalization group, quantum master equations, Keldysh Green's function technique and dynamical mean-field theory \cite{Schollwock2005,Kamenev2009,Breuer2002}, we still lack a schematic framework to interpret the vast kinds of nonequilibrium phenomena.

Probability distribution of quantum work is arguably one of the most important quantities to characterize the nonequilibrium dynamics such as the dynamical fluctuations, quantum phase transitions (QPTs) and quantum criticality
\cite{Imparato2005,Engel2007,Palmai2014,Fusco2014,Talkner2016,Lobejko2017,Modak2017,Wang2017,Funo2017,Russomanno2015,Shraddha2015,Jarzynski2015,Nigro2018}. Technically, it may be defined as
\begin{equation}
p(w)  =\sum_{n,m}\delta(W-E_{m}^{f}+E_{n}^{i})P\left(  m^{f}\mid
n^{i}\right)  P\left(  n^{i}\right)  , \label{wdf}%
\end{equation}
where $w=W/N^d$ is the work density, $N$ is the lattice size, $d$ is the dimensionality, $E_{n}^{i}$ is the $n$-th eigenvalue of the initial state ($|n^i\rangle$), and $E_{m}^{f}$ is the $m$-th eigenvalue of the final state ($|m^f\rangle$). $P(  n^{i})=\exp(  -\beta E_{n}^{i})  /Z(  0)  $ denotes the probability distribution of the initial canonical state, and $P(  m^{f}\mid n^{i})\equiv\vert \langle m^{f}\vert U(  T_{0},0)\vert n^{i}\rangle \vert ^{2}$ accounts for the transition probability between the initial and final states governed by the time-dependent evolution, $U(  T_{0},0)  =\mathcal{T}\exp\left[  -\operatorname*{i}\int_{0}^{T_{0}}dtH(  t)  \right]  $, where $\mathcal{T}$ is the time ordering operator and $T_0$ is the period for the time evolution. The reduced Planck constant $\hbar$ is set to unity. The nonequilibrium dynamics is fully incorporated in the time-dependent Hamiltonian $H(t)$ and the transition probability contains the key information characterizing the dynamical process.

The above method has been applied to a variety of many-body systems including the transverse Ising chain \cite{Silva2008,Dorner2012}, the anisotropic XY model \cite{Bayocboc2015}, the XXZ model \cite{Mascarenhas2014}, the Luttinger liquid \cite{Dora2012}, and the low-dimensional quantum gas \cite{Gambassi2012,Sotiriadis2013,Shchadilova2014}. In most of these works, a Hamiltonian approach has been used by calculating explicitly the time evolution of the quantum state for quench protocol. However, such calculations can be very involved which makes it hard to be extended to arbitrary time dependence and general correlated many-body systems \cite{Dora2012,Smacchia2013}. To overcome this issue, here we propose to carry out the calculations straightforwardly using the functional field integral formalism, which is a many-body extension of the path integral approach to quantum fields \cite{Altland2010}. Our approach has the advantage of numerical simplicity and may be extended to more complicated nonequilibrium processes beyond the quench protocol. Moreover, in combination with the auxiliary field method and mean-field approximation, it has the potential to be applied to more general correlated systems, in the cost of explicit state evolution.

The manuscript is organized as follows. In Section \ref{formalism}, we first introduce the general formalism based on the functional field integral approach and show that it can be used to identify both the equilibrium and dynamical QPTs. In Section \ref{model}, we apply it to three different models, including the well-studied transverse Ising model, the Su-Schrieffer-Heeger (SSH) model with topological phase transition and the Bardeen-Cooper-Schrieffer (BCS) model with superconductivity under mean-field approximation. We will discuss the QPTs in these models and derive the corresponding quantum critical exponents from the calculated mean irreversible work density. Section \ref{con} is the discussion and conclusions.

\section{General formalism}
\label{formalism}

\begin{figure}[b]
\begin{center}
\includegraphics[width=8.5cm]{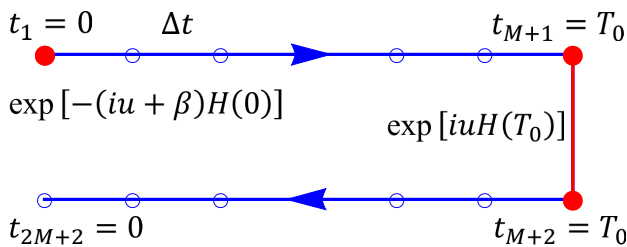}
\end{center}
\caption{(Color online) Time contour $\mathcal{C}$ for the field integral. Blue circles represent discretized time points with $\Delta t=T_0/M$. Red dots and line at $t=0$ and $T_0$ denote additional operators to be evaluated at these points.}
\label{fig1}
\end{figure}

To proceed, we consider the bilinear model,
\begin{equation}
H\left(  t\right)  =\sum_{k}\Psi_{k}^{\dagger}A_{k}\left(  t\right)  \Psi_{k},
\label{BdG}
\end{equation}
where $A_{k}(  t)  =\mathbf{d}_{k}(  t)  \cdot\boldsymbol{\sigma}$ is a matrix and $\boldsymbol{\sigma}$ is the vector of the Pauli matrices. Although simple, it represents a large number of models in condensed matter physics. We first reformulate the functional field integral approach to calculate the Fourier transformation of the work distribution, namely the characteristic function \cite{Talkner2007PRE}:
\begin{align}
&  G(u)  =\int_{-\infty}^{\infty}dw\exp\left(  \operatorname*{i}uwN^d\right)  p(w) \nonumber\\
&  =Z_{0}^{-1}\operatorname*{Tr}\left[  U^{\dagger}\left(  T_{0},0\right)\operatorname{e}^{\operatorname*{i}uH\left(  T_{0}\right)  }U\left(
T_{0},0\right)  \operatorname{e}^{-\left(  \operatorname*{i}u+\beta\right)H\left(  0\right)  }\right]  , \label{gf}
\end{align}
where $Z_{0}=\operatorname*{Tr}  \operatorname{e}^{-\beta H\left(0\right)  }  $ is the partition function at $t=0$. The time contour of the path  integral in $G(u)$ is illustrated in Fig.~\ref{fig1}. After some tedious calculations using the functional field integral techniques (see Appendix \ref{appA}), we obtain
\begin{equation}
G(u)  ={\displaystyle\prod\limits_{k}}
\frac{2+\operatorname*{Tr}\left[  B_{k}\left(  T_{0}\right)  \right]
}{2+\operatorname*{Tr}\left[  \operatorname{e}^{-\beta A_{k}\left(  0\right)
}\right]  }, \label{Gu}
\end{equation}
where
\begin{equation}
B_{k}(T_{0})=C_{k}^{\dagger}\left(  T_{0}\right)  \operatorname{e}^{\operatorname*{i}uA_{k}\left(  T_{0}\right)  }C_{k}\left(  T_{0}\right)
\operatorname{e}^{-\left(  \operatorname*{i}u+\beta\right)  A_{k}\left(0\right)  },
\label{Bk}
\end{equation}
with $C_{k}\left(  T_{0}\right)  =\mathcal{T}\exp\left[  -\operatorname*{i}\int_{0}^{T_{0}}dtA_{k}\left(  t\right)  \right]  $. The mean work density is given by the first cumulant of the characteristic function, $\left\langle w\right\rangle =-\operatorname*{i}dG(u)  /(N^ddu)|_{u=0}$, yielding
\begin{equation}
\left\langle w\right\rangle =\frac{1}{N^d}\sum_{k}\frac{\operatorname*{Tr}\left[  \left(
D_{k}\left(  T_{0}\right)  -A_{k}\left(  0\right)  \right)  \operatorname{e}^{-\beta A_{k}\left(  0\right)  }\right]  }{2+\operatorname*{Tr}\left[
\operatorname{e}^{-\beta A_{k}\left(  0\right)  }\right]  },
\end{equation}
with $D_{k}(  T_{0})  =C_{k}^{\dagger}(  T_{0})A_{k}(  T_{0})  C_{k}(  T_{0})  $, where $C_{k}(T_{0})$ can be computed numerically for arbitrary time dependence.

Note that the mean work density $\left\langle w\right\rangle $ always exceeds the free energy density difference $\Delta f$ between the initial and final equilibrium states (both with the same inverse temperature $\beta$), as stated in the second law of thermodynamics. A mean irreversible work density can thus be defined as their difference, $\langle w_{\text{irr}}\rangle =\langle w\rangle -\Delta f\geq0$, which characterizes the irreversibility of the nonequilibrium process and is directly related to the entropy increase between the final and initial equilibrium states, $\Delta s=\beta\langle w_{\text{irr}}\rangle $, for a closed quantum system without heat transfer \cite{Shraddha2015}.

The mean work density $\left\langle w\right\rangle $ and the mean irreversible work density $\left\langle w_{\text{irr}}\right\rangle $ can be used to identify equilibrium phase transitions. Considering a quench process where the model Hamiltonian changes from $A_{k}^{0}=\mathbf{d}_{k}^{0}\cdot \boldsymbol{\sigma}$ at time $t=0^{-}$ to $A_{k}^{1}=\mathbf{d}_{k}^{1}\cdot\boldsymbol{\sigma}$ at $t=T_{0}=0^{+}$, we have $D_{k}(0^{+})  =A_{k}^{1}$. As shown in Appendix \ref{appB}, one can derive an explicit formula for the free energy density difference,
\begin{equation}
\Delta f=-\frac{1}{\beta N^d}\sum_{k}\ln\frac{\cosh^{2}\left(  \beta E_{k}^{1}/2\right)  }{\cosh^{2}\left(  \beta E_{k}^{0}/2\right)  },
\label{freeenergy}
\end{equation}
and the mean work density,
\begin{equation}
\left\langle w\right\rangle =\frac{1}{N^d}\sum_{k}\frac{\left[  \left(  E_{k}^{0}\right)^{2}-\mathbf{d}_{k}^{0}\cdot\mathbf{d}_{k}^{1}\right]  \tanh\left(  \frac
{1}{2}\beta E_{k}^{0}\right)  }{E_{k}^{0}},
\label{meanwork}
\end{equation}
where $E_{k}^{0,1}=\left\vert \mathbf{d}_{k}^{0,1}\right\vert $. Then the mean irreversible work density can be immediately obtained by definition. To see how these detect the phase transitions, we consider a small quench, $\mathbf{d}_{k}^{0}=\left(  x_{k}^{0},y_{k}^{0},z_{k}^{0}\right)\rightarrow\mathbf{d}_{k}^{1}=\left(  x_{k}^{0}+\delta,y_{k}^{0},z_{k}^{0}\right) $. It can be shown analytically (Appendix B) that $\left\langle w\right\rangle=-\sum_{k}\frac{\delta}{N^d}\frac{\partial E_{k}}{\partial x_{k}}|_{x_{k}=x_{k}^{0}}$ and $\left\langle w_{\text{irr}}\right\rangle =\sum_{k}\frac{\delta^{2}}{2N^d}\frac{\partial^{2}E_{k}}{\partial x_{k}^{2}}|_{x_{k}=x_{k}^{0}}$ at zero temperature, which connect directly to the first and second derivatives of the ground state energy with respect to the quench parameter. Thus the singularity in $\left\langle w\right\rangle $ and $\left\langle w_{\text{irr}}\right\rangle $ reflect the first or second-order phase transitions.

For a (second-order) QPT, scaling analysis of $\left\langle w_{\text{irr}}\right\rangle $ can yield key information on the critical exponents \cite{Shraddha2015}. In the heat susceptibility limit, where $\delta^{-v}$ is the largest length scale, the mean irreversible work density at zero
temperature scales as
\begin{align}
\left\langle w_{\text{irr}}\right\rangle/\delta^{2}  &  \sim\lambda^{\nu\left(  d+z\right)  -2},\ \ \quad\delta^{-\nu}>N>\lambda^{-\nu
}\nonumber\\
&  \sim N^{2/\nu-\left(  d+z\right)  },\quad\delta^{-\nu}>\lambda^{-\nu}>N,
\end{align}
where $\nu$ is the correlation length exponent, $z$ is the dynamical exponent, and $\lambda$ reflects the distance from the quantum critical point (QCP). In the thermodynamic limit, where $N$ is the largest length scale, one arrives at the scaling relation
\begin{align}
\left\langle w_{\text{irr}}\right\rangle/\delta^{2}   &  \sim\delta^{\nu\left(d+z\right)-2  },\quad\quad\quad N>\lambda^{-\nu}>\delta^{-\nu}\nonumber\\
&  \sim\lambda^{\nu\left(  d+z\right)  -2},\quad\quad\quad N>\delta^{-\nu}>\lambda^{-\nu}.
\end{align}
On the other hand, if the system is prepared very close to the QCP such that $\lambda^{-\nu}$ is larger than all other length scales, the scaling relation becomes
\begin{align}
\left\langle w_{\text{irr}}\right\rangle/\delta^{2}   &  \sim\delta^{\nu\left(d+z\right)-2},\quad\quad\quad\ \ \lambda^{-\nu}>N>\delta^{-\nu}\nonumber\\
&  \sim N^{2/\nu-\left(  d+z\right)  },\quad\quad\ \ \,\lambda^{-\nu}>\delta^{-\nu}>N.
\end{align}
Note that when the combination $\nu\left(  d+z\right)  $ exceeds 2, the scaling of $\left\langle w_{\text{irr}}\right\rangle $ is non-universal. In the marginal case, $\nu\left(  d+z\right)  =2$, one may find additional logarithmic correction to above scalings \cite{Shraddha2015}.

Recently, it has also been shown that the work statistics in a double quench process, where the Hamiltonian changes from $A_{k}^{0}$ to $A_{k}^{1}$ at $t=0$ and then back to $A_{k}^{0}$ for $t\ge T_{0}$, can be used to describe dynamical QPTs \cite{Heyl2013}. In this case, the free energy density difference $\Delta f$ is zero and we have $\langle w_{\text{irr}}\rangle=\langle w\rangle $ and $D_{k}( T_{0}) =e^{ \operatorname*{i}T_{0}A_{k}^{1}} A_{k}^{0}e^{ -\operatorname*{i}T_{0}A_{k}^{1}}$. The mean irreversible work density is incapable of capturing the dynamical QPTs. On the other hand, the work distribution function, which in principle could be evaluated by inverse Fourier transformation of $G(u)$, is often ill-defined in numerical calculations. Lately, an alternative approach has been proposed based on the so-called G\"{a}rter-Ellis theorem \cite{Abeling2016,Majumdar2017}. Considering a global quench process, the
quantum work grows exponentially with the system size, $p(w)  \sim e^{-N^dr(w)  }$, which defines the rate function $r(w)  \ge 0$. The theorem states that $r(w)$ can be obtained via a Legendre-Fenchel
transformation,
\begin{equation}
r( w) =-\inf_{R\in\mathbb{R}}[ wR-c( R) ], \label{rw}
\end{equation}
where $c( R) =-\lim_{N\rightarrow\infty}N^{-d}\ln G(u=\operatorname*{i}R)$ is a scaled cumulant generating function assumed to be differentiable with respect to the real variable $R\in\mathbb{R}$, and the infimum is evaluated within the domain of definition of $c(R) $ including $R=\pm\infty$. The dynamical QPT is then manifested as a singular point of $r(w)$, as discussion in Appendix \ref{appB}.

For general time dependence, $C_{k}(T_{0})$ may be computed numerically using the matrix products,
\begin{equation}
C_{k}\left(  T_{0}\right)  \approx \operatorname{e}^{-\operatorname*{i}\Delta tA_{k}\left(  \frac{t_{M+1}+t_{M}}{2}\right)  }\cdots\operatorname{e}
^{-\operatorname*{i}\Delta tA_{k}\left(  \frac{t_{1}+t_{2}}{2}\right)  },
\end{equation}
which becomes accurate as $M\rightarrow \infty$ and $\Delta t=T_0/M \rightarrow 0$. If all $A_k(t)$ commute, the above formula reduces to a simple time integral of $A_k(t)$ in the exponent. It is anticipated that detailed investigations for arbitrary time dependence might reveal more interesting properties of nonequilibrium physics. In this work, we will focus on quench protocols for the sake of simplicity and only give as an example a brief discussion on general cases in Section \ref{ssh}.

\section{Numerical results}
\label{model}

In this section, we use the above general formalism to study the work statistics and possible QPTs in three different models. The first model is the transverse Ising model, which has been extensively studied using the Hamiltonian approach and hence provides a good examination of our method. We then discuss the SSH model, where we will find corrections to the quantum work due to its topological properties. Last but not least, for possible extension to other correlated models in future studies, we discuss the well-known BCS model for superconductivity in combination with the mean-field approximation.

\subsection{The transverse Ising model}

The Hamiltonian of the transverse Ising model is \cite{Pfeuty1970,Suzuki2013,Dutta2015}
\begin{equation}
H=-J\sum_{j=1}^{N}\sigma_{j}^{x}\sigma_{j+1}^{x}-h\sum_{j=1}^{N}\sigma_{j}^{z},
\label{Hisng}
\end{equation}
where $\sigma_{j}^{\alpha}\,(  \alpha=x,z)  $ are the Pauli matrices at site $j$, $J>0$ is the ferromagnetic exchange coupling, $h$ is the transverse external field. Here we assume $N$ is even and consider the periodic boundary condition, $\sigma_{N+1}^{\alpha}=\sigma_{1}^{\alpha}$. This model has a quantum critical point at $h=J$. Using the Jordan-Wigner transformation,
\begin{equation}
\sigma_{j}^{+}=\frac{\left(  \sigma_{j}^{x}+\operatorname*{i}\sigma_{j}^{y}\right)  }{2}=c_{j}^{\dagger}\exp\left(  \operatorname*{i}\pi\sum
_{l<j}c_{l}^{\dagger}c_{l}\right)  ,
\end{equation}
it can be mapped to a spinless fermion model \cite{Dziarmaga2005,Franchini2017,Dong2018},
\begin{equation}
H=Nh-2h\sum_{j=1}^{N}c_{j}^{\dagger}c_{j}-J\sum_{j=1}^{N}\left(c_{j}^{\dagger}-c_{j}\right)  \left(  c_{j+1}^{\dagger}-c_{j+1}\right)  ,
\end{equation}
where the fermionic operator $c_{j}$ satisfies either periodic or anti-periodic boundary conditions, $c_{N+1}=\pm c_{1}$, depending on odd or even number of the $c$-quasiparticles, $M=\sum_{j=1}^{N}c_{j}^{\dagger}c_{j}$. For simplicity, we confine ourselves to the subspace of even $M$ with anti-periodic boundary condition. Using $c_{q}=N^{-1/2}\sum_{j=1}^{N}c_{j}\exp(\operatorname*{i}qj)$, the fermionic Hamiltonian can be transformed into the bilinear form, $H=\sum_{k>0}\Psi_{k}^{\dagger}\mathbf{d}_{k} \cdot\boldsymbol{\sigma}\Psi_{k}$, where
$\Psi_{k}^{\dagger}=(c_{k}^{\dagger},c_{-k})$, $\mathbf{d}_{k}=(0,-2J\sin k,-2h-2J\cos k)$, and $k=\pm\pi(2m-1)/N$ with $m=1,\ldots,\,N/2$.

\begin{figure}[pt]
\begin{center}
\includegraphics[width=8.6cm]{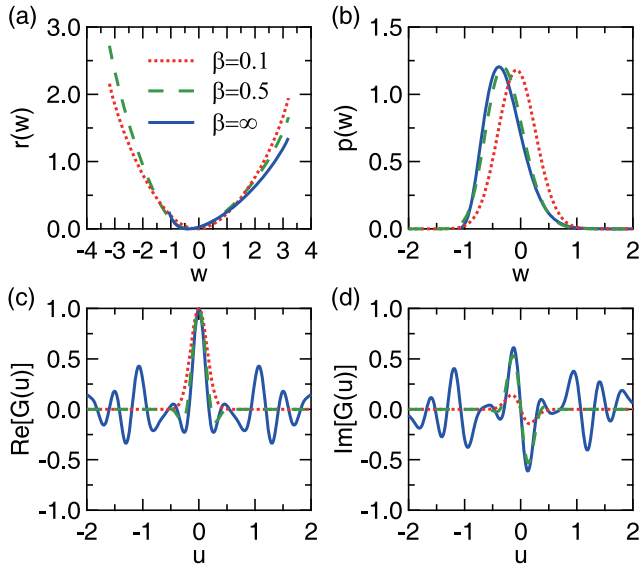}
\end{center}
\caption{(Color online) (a) The rate function $r(w)$ for a single quench from $h_{0}=0.5$ to $h_{1}=2.0$ in the transverse Ising model at different inverse temperature $\beta$. (b) The corresponding work distribution function $p(w)\sim e^{-Nr(w)  }$. (c) and (d) are the real and imaginary parts of the characteristic function $G(u)  $, respectively. The lattice size is set to $N=20$.}
\label{figisingrate}
\end{figure}

\begin{figure}[pt]
\begin{center}
\includegraphics[width=8.6cm]{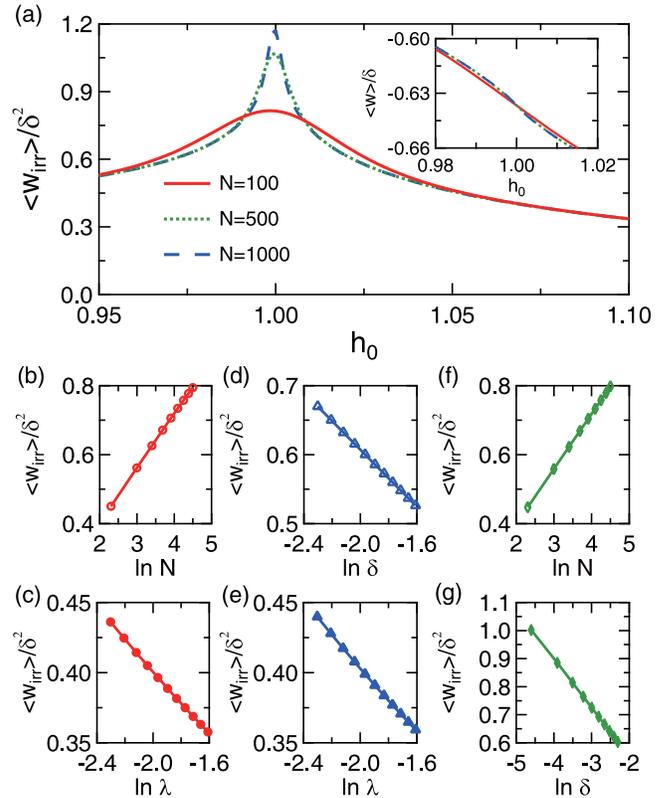}
\end{center}
\caption{(Color online) (a) The mean irreversible work density, $\langle w_{\text{irr}}\rangle/\delta^{2}$, as a function of $h_{0}$ with different lattice size $N$. The inset shows the $h_0$ dependence of the mean work density $\langle w\rangle /\delta $. (b) and (c) show the logarithmic dependence of $\langle w_{\text{irr}}\rangle/\delta^{2}$ on $N$ and $\lambda$ in the heat susceptibility limit ($\delta=0.001$). Other parameters are $\lambda=0.005$ in (b)\ and $N=100$ in (c). (d) and (e) show its logarithmic scaling with respect to $\delta$ and $\lambda$ in the thermodynamic limit ($N=1000$) for $\lambda=0.01$ in (d) and $\delta=0.01$ in (e). (f) and (g) show the logarithmic scaling with $N$ and $\delta$ for $\lambda=0.0005$, such that $\lambda^{-\nu}$ is the largest length scale and the prequench Hamiltonian is very close to the quantum critical point. Other parameters are $\delta=0.001$ in (f)\ and $N=1000$ in (g).}
\label{figIsing-scaling}
\end{figure}

\begin{figure}[ptb]
\begin{center}
\includegraphics[width=8.6cm]{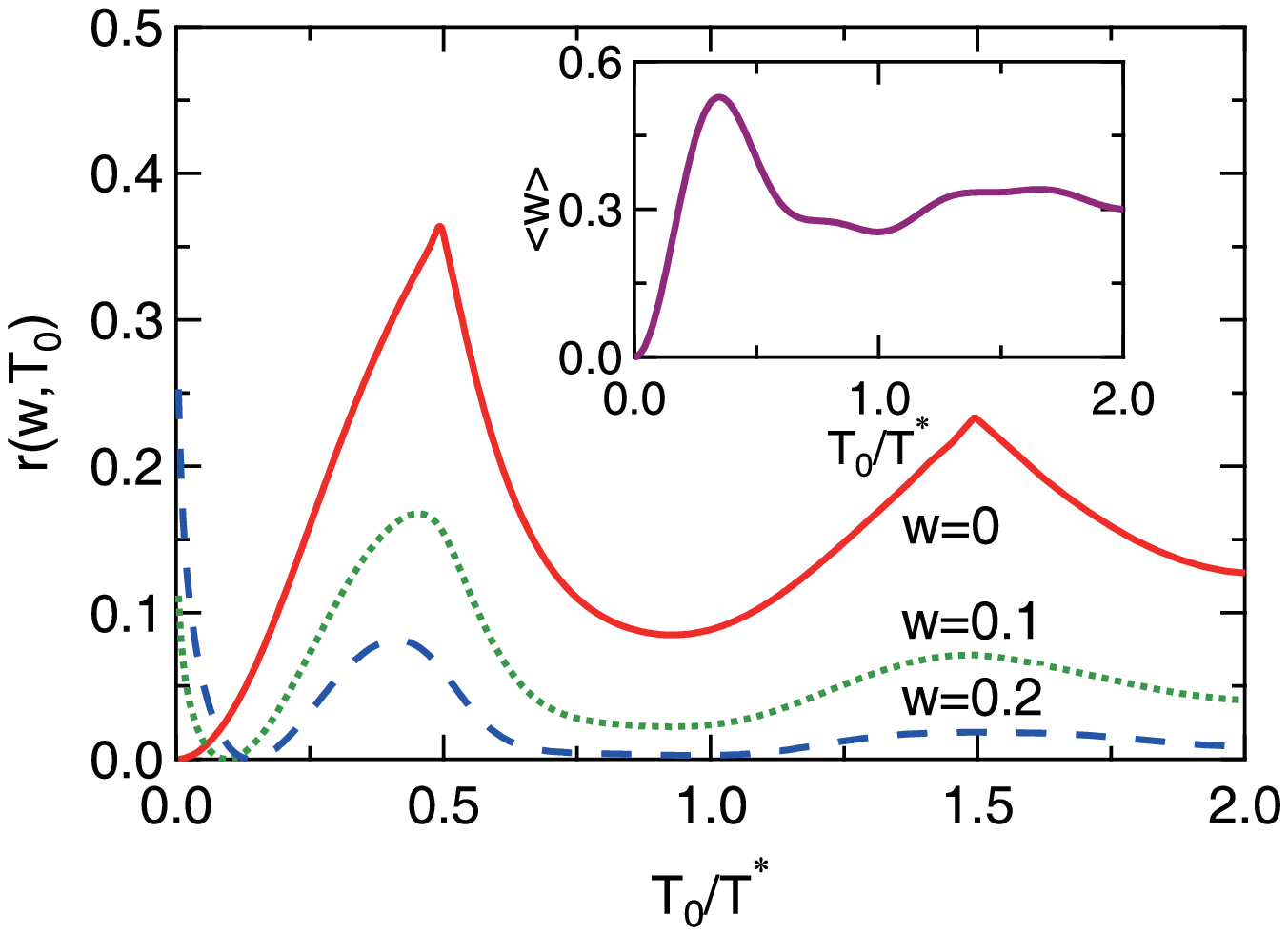}
\end{center}
\caption{(Color online) The rate function $r(  w,T_{0})  $ at zero temperature for a double quench process of the transverse Ising model from $h_{0}=0.5$ to $h_{1}=2.0$ and back to $h_{0}$ after time $T_{0}$. The lattice size is $N=100$. The rate function for $w=0$ corresponds to the Loschmidt echo and its nonanalyticity at $T_0=(  n+1/2)  T^{\ast}$ manifests the dynamical QPTs. The inset shows the mean work density $\langle w\rangle $, which exhibits no singularity and is a smooth function of $T_0$.}
\label{figIsingperiodic}
\end{figure}

Below we set $J$ to unity and consider the quench protocol where the external field $h$ is tuned from its initial value $h_{0}$ to $h_{1}=h_{0}+\delta$. The dispersion relation is $\epsilon_{k}(  h)=\vert \mathbf{d}_{k}\vert =2\sqrt{1+h^{2}+2h\cos k}.$ The characteristic function $G(u)  =\prod_{k>0}G_{k}(  u)$ for transverse Ising chain has an analytical form as shown in Eq.
(\ref{charageneral}) of Appendix \ref{appB}. Figure \ref{figisingrate} plots the characteristic function for a quench protocol across the QCP and the corresponding work distribution function obtained through the G\"{a}rter-Ellis theorem at different temperatures.
As discussed in Appendix \ref{appB}, the work distribution is restricted in the interval, $[  w_{\min},w_{\max}]$, where $w_{\min}$ is the energy density difference of the highest excited state of the initial phase and the ground state of the final phase and $w_{\max}$ is the energy density difference between the ground state of the initial phase and highest excited state of the final phase. At finite temperature, the small but finite $p(w)$ close to $w_{\min}$ and $w_{\max}$ reflects the finite weight of all possible configurations due to thermal effect. However, at zero temperature, since there is no weight for the excited state in the initial phase, $p(w)$ is strictly zero if $w$ is smaller than the energy density difference of the ground states of the initial and final Hamiltonians, as can be seen in Fig.~\ref{figisingrate}(b). For all the temperatures, we have the normalization condition, $G(  0)  =\int_{-\infty}^{\infty}dw\,p(w)  =1$, as confirmed in Figs. \ref{figisingrate}(c) and (d). These results agree with those obtained using the Hamiltonian eigenstate approach \cite{Abeling2016}.

The mean work density and mean irreversible work density at zero temperature can be obtained from Eqs.~(\ref{freeenergy}) and (\ref{meanwork}). Figure  \ref{figIsing-scaling}(a) plots the variation of $\langle w\rangle /\delta $ and $\langle w_{\text{irr}}\rangle /\delta^{2}$ as a function of $h_{0}$ with different lattice size $N$. The mean work density changes continuously with the external field $h_{0}$ and shows no visible features across the QCP, while a significant peak is seen to grow with increasing $N$ in the mean irreversible work density, indicating that the latter is a good indicator of the second order QPT as discussed in Section \ref{formalism}. The quantum criticality can then be  examined from the scaling behavior in different parameter regimes. In Fig. \ref{figIsing-scaling}, we see clear logarithmic corrections in $\langle w_{\text{irr}}\rangle $ depending on $N$, $\lambda$, or $\delta$, in agreement with the critical exponent $\nu=z=1$ of the transverse Ising chain.

We now turn to the double quench protocol where the external field $h$ is tuned from $h_{0}<1$ to $h_{1}>1$ at time $t=0$ and quenched back to $h_{0}$ at $t=T_{0}$. Figure \ref{figIsingperiodic} plots the rate function as a function of $T_0$ for different values of the work density $w=W/N$. We see clear cusp in the curve of $w=0$ and continuous variation for other $w$. Such nonanalytic behavior at $w=0$ in the rate function is an indication of the so-called dynamical QPT, which occurs when the time-evolving state $\vert\psi(  t)  \rangle $ is orthogonal to the initial state at certain critical time after quenching a set of control parameters of the Hamiltonian. As discussed in Appendix \ref{appC} for the transverse Ising model, there exists a sequence of critical time, $t_{c}=(n+1/2)  T^{\ast}$, with
\begin{equation}
T^{\ast}=\frac{\pi}{2\sqrt{1+\left(  h_{1}\right)  ^{2}-2h_{1}\frac
{1+h_{0}h_{1}}{h_{0}+h_{1}}}}.
\end{equation}
Its inverse, $\omega^\ast=2\pi/T^{\ast}$, seems to define a characteristic excitation energy for the transition \cite{Piccitto2019}.
We should note that such singularity is not present in the mean (irreversible) work density. As shown in the inset, the mean work density varies smoothly with $T_0$ and is insensitive to the dynamical QPT revealed in $r(w, T_0)$. The above results are in good agreement with previous studies using the Loschmidt echo \cite{Heyl2013}, which is defined as $L\left(  t\right)  =\left\vert \left\langle \psi\left\vert U\left(  t\right)\right\vert \psi\right\rangle \right\vert ^{2} $ and represents the probability amplitude to recover the initial state after the time evolution $U\left(  t\right)  $. For double quench process, the work probability function $p\left(  w=0,T_{0}\right)  $ also gives the return probability to the initial state and could therefore provide a signature when a dynamical QPT occurs \cite{Abeling2016}. The excellent agreement between our results and previous Hamiltonian eigenstate method confirms the validity of our approach for both the single and double quench protocols. We may extend it to other models and examine there the effect of quantum criticality on nonequilibrium dynamics and the possible existence of dynamical QPTs under more general circumstances.

\subsection{The SSH model}

\label{ssh}

\begin{figure}[ptb]
\begin{center}
\includegraphics[width=8.6cm]{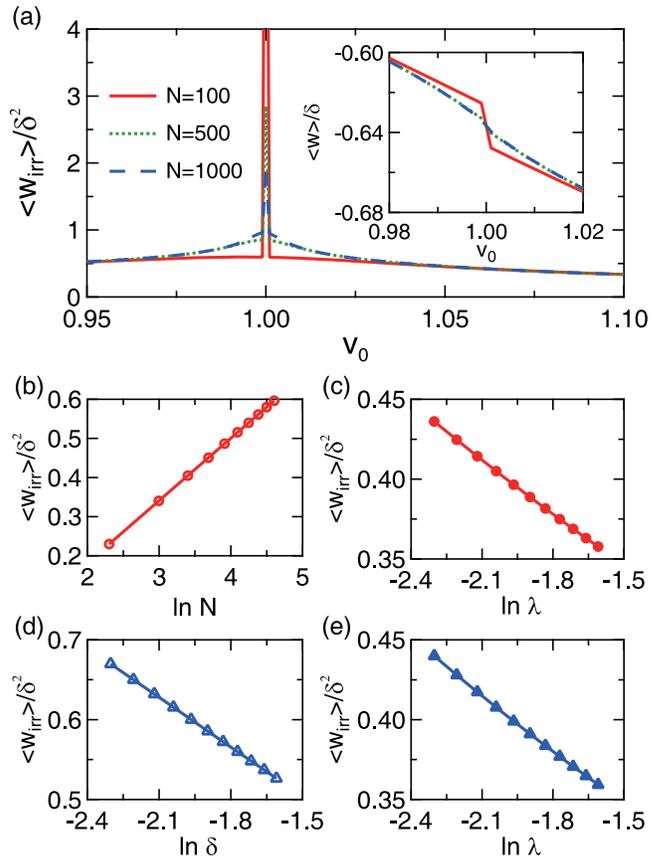}
\end{center}
\caption{(Color online) (a) The mean irreversible work density, $\langle w_{\text{irr}}\rangle/\delta^{2}$, as a function of $v_{0}$ for different lattice size $N$. The inset plots the mean work density $\langle w\rangle /\delta $ versus $v_0$, showing a clear jump at $v_0=1$ for small $N$. (b) and (c) show the logarithmic dependence of $\langle w_{\text{irr}}\rangle/\delta^{2}$ on $N$ and $\lambda$ in the heat susceptibility limit ($\delta=0.001$) for $\lambda=0.005$ in (b) and $N=100$ in (c). (d) and (e) show its logarithmic scaling with respect to $\delta$ and $\lambda$ in the thermodynamic limit ($N=1000$) for $\lambda=0.01$ in (d) and $\delta=0.01$ in (e).}
\label{figsshN}
\end{figure}

In this section, we study the SSH model which was originally proposed for electronic transport in polyacetylene with spontaneous dimerization \cite{Su1979}. Despite its simplicity, the SSH model exhibits a variety of exotic phenomena, such as topological soliton excitation, fractional charge and nontrivial edge states, and has attracted extensive interest in past decades \cite{Li2014,Shen2012,Bernevig2013,Asboth2016,He2016}. The model Hamiltonian can be written as
\begin{equation}
H=\sum_{j=1}^N\left(  vc_{A,j}^{\dagger}c_{B,j}+ v^{\prime} c_{A,j+1}^{\dagger}c_{B,j}+h.c.\right)  ,
\end{equation}
where $A$ and $B$ denote the two sublattices and $N$ is the number of unit cells. For even $N$ and open boundary condition, there exist two edge modes for $v<v^{\prime}$ but no edge mode for $v>v^{\prime}$. Thus the model undergoes a topological quantum phase transition at $v=v^{\prime}$. For periodic boundary condition, one may apply the Fourier transformation, $c_{\alpha,j}=N^{-1/2}\sum_{k}c_{\alpha,k}\exp\left(  \operatorname*{i}kj\right)  $, where $\alpha=A$ or $B$, $k=2m\pi/N$ with $m=1-N/2,\ldots,\,N/2$. Then the bulk Hamiltonian gets the bilinear form, $H=\sum_{k}\Psi_{k}^{\dagger}\mathbf{d}_{k} \cdot\boldsymbol{\sigma}\Psi_{k},$ where $\Psi_{k}^{\dagger}=\left(  c_{A,k}^{\dagger},c_{B,k}^{\dagger}\right)  $ and $\mathbf{d}_{k}=\left(  v+\cos k, \sin k,0\right)  $.

\begin{figure}[t]
\begin{center}
\includegraphics[width=8.6cm]{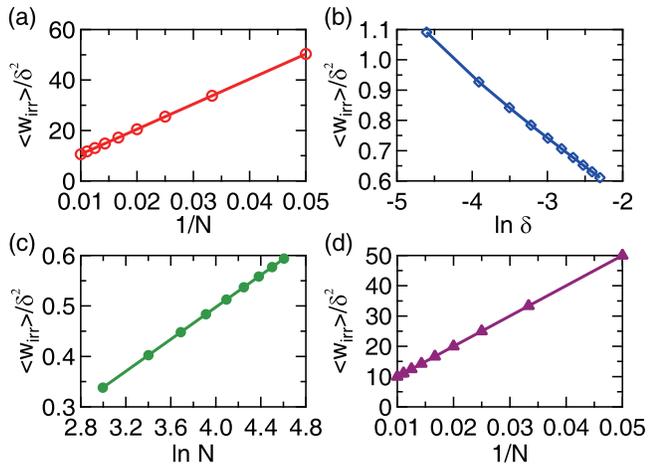}
\end{center}
\caption{(Color online) (a) and (b) show the scaling behavior of the mean irreversible work density, $\langle w_{\text{irr}}\rangle/\delta^{2}$, with $N$ in the SSH model near the quantum critical point. We take $\lambda=0.0005$ such that $\lambda^{-\nu}$ is the largest length scale of the system. Other parameters are $\delta=0.001$ in (a) and $N=1000$ in (b). The $1/N$ scaling in (a) can be separated into (c) a small $\ln N$ contribution from quantum criticality and (d) a dominant nonuniversal $1/N$ contribution due to the topological nature of the QPT in the SSH model.}
\label{figsshN2}
\end{figure}

Details on the calculations of the mean work density and the mean irreversible work density can be found in the Appendix \ref{appC}. Figure \ref{figsshN}(a) plots the results as a function of $v_0$ for a single quench from $v=v_0$ to $v_0+\delta$ at $v^{\prime}=1$. In contrast to that of the transverse Ising model shown in Fig. \ref{figIsing-scaling}, the mean work density exhibits a clear discontinuity at the critical point. Correspondingly, one observes a sharp resonance-like peak in the mean irreversible work density on a smooth background. As $N$ increases, the background evolves into a broad peak, resembling that in the transverse Ising model, but the sharp resonance at the critical point becomes weakened. Except for a small region of the sharp peak around critical $v_0=1$, the mean irreversible work density in all other parameter ranges from the relatively smooth background exhibits similar logarithmic scaling with respect to $N$, $\lambda$ or $\delta$. This is plotted in Figs. \ref{figsshN}(b-e) and indicates that the critical exponents are $\nu=z=1$, as in the transverse Ising model. In contrast, as one may see from Figs.~\ref{figsshN2}(a) and (b), the mean irreversible work density deviates from the expected scaling at $v_0=1$ ($\lambda=0$) due to the presence of the sharp resonance. As a matter of fact, we find when $\lambda^{-\nu}$ is the largest length scale, the scaling relation becomes
\begin{align}
\frac{\left\langle w_{\text{irr}}\right\rangle}{\delta^{2}} &  \sim \ln\delta+\frac{2}{N}\frac{1}{\delta}-\frac{2\lambda}{N}\frac{1}{\delta^2},\quad
\lambda^{-\nu}>N>\delta^{-\nu}\nonumber\\
&  \sim\ln N+\frac{2\left(  \delta-\lambda\right)  }{\delta^{2}}\frac{1}{N},\quad\lambda^{-\nu}>\delta^{-\nu}>N.
\end{align}
Such non-universal $1/N$ contributions are associated with the topological nature of the SSH model. A straightforward analysis (Appendix \ref{appC}) suggests that they originate from the band crossing at $k=\pi$ at the topological quantum phase transition. One may separate the contribution from this peculiar point and obtain,
\begin{equation}
\frac{\left\langle w_{\text{irr}}\right\rangle _{k=\pi}}{\delta^{2}}%
=\frac{2\left(  \delta-\lambda\right)  }{\delta^{2}}\frac{1}{N}.
\end{equation}
For $\lambda^{-\nu}>\delta^{-\nu}>N$ and $\nu=1$, in particular, this $1/N$ term has large coefficient and overwhelms the $\ln N$ term as shown in Fig. \ref{figsshN2}(a). To see this more clearly, we plot the mean irreversible work density from momenta $k\neq\pi$ and $k=\pi$ in Figs. \ref{figsshN2}(c) and (d), respectively. Indeed, there exists a clear and small logarithmic contribution, which is of the same order of magnitude as in Fig.~\ref{figIsing-scaling}(b), but a huge $1/N$ contribution.  Intuitively, when the Hamiltonian is quenched across the topological QCP, the necessary existence of the band crossing yields this $1/N$ term. For open boundary condition, these correspond to the  edge modes whose contribution to the mean irreversible work density scales inversely with the lattice size $N$.

\begin{figure}[t]
\begin{center}
\includegraphics[width=8.6cm]{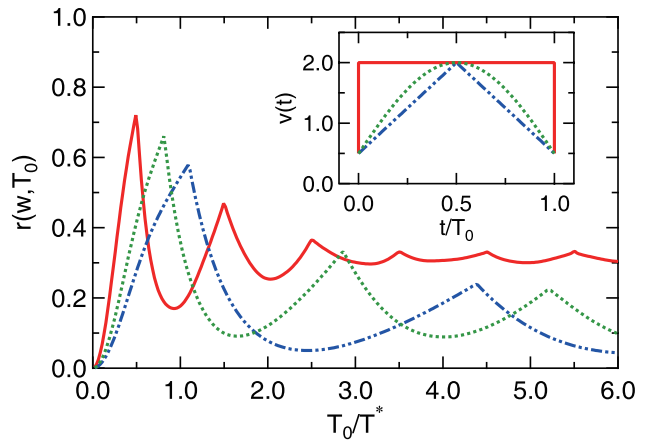}
\end{center}
\caption{(Color online) The rate function $r(w,T_0)$ at $w=0$ and zero temperature for three different time evolution protocols of the SSH model. The inset plots the corresponding $v(t)$ with $T_0=2.0$. The lattice size is $N=1000$. The singularities in the rate function reflect the dynamical QPTs in all three cases.}
\label{figSSH-DQPT}
\end{figure}

The rate function for the double quench protocol of the SSH model is plotted in Fig. \ref{figSSH-DQPT} and actually similar to that of the Ising model (up to a factor of 2) since both models have the same bilinear Hamiltonian except for the boundary conditions. Thus their difference is delicate and actually diminishes in the thermodynamic limit. For comparison, we also plot the results for two other protocols with different time dependent $v(t)$ as shown in the inset. Interestingly, we see clear cusps in $r(w,T_0)$ at $w=0$ and zero temperature in all three cases, indicating the possibility of general existence of dynamical QPTs. However, no time periodicity is seen except for the double quench case, where the critical time is given by $t_c=\left(  n+1/2\right)  T^{\ast}$ with
\begin{equation}
T^{\ast}=\frac{\pi}{\sqrt{1+\left(  v_{1}\right)  ^{2}-2v_{1}\frac
{1+v_{0}v_{1}}{v_{0}+v_{1}}}}.
\end{equation}
More elaborate investigations might be able to reveal the true controlling parameter for the dynamical QPT. We note that since the dynamical QPT in the SSH model is associated with the equilibrium QPT with topological properties, it might be intriguing to think if the nonequilibrium phase might also consist in some topological properties, for example, dynamic edge modes under open boundary conditions \cite{Y-Wang2017PRB,Y-Wang2018PRE}.

\subsection{The BCS model}
\label{BCS}

\begin{figure}[pt]
\begin{center}
\includegraphics[width=8.6cm]{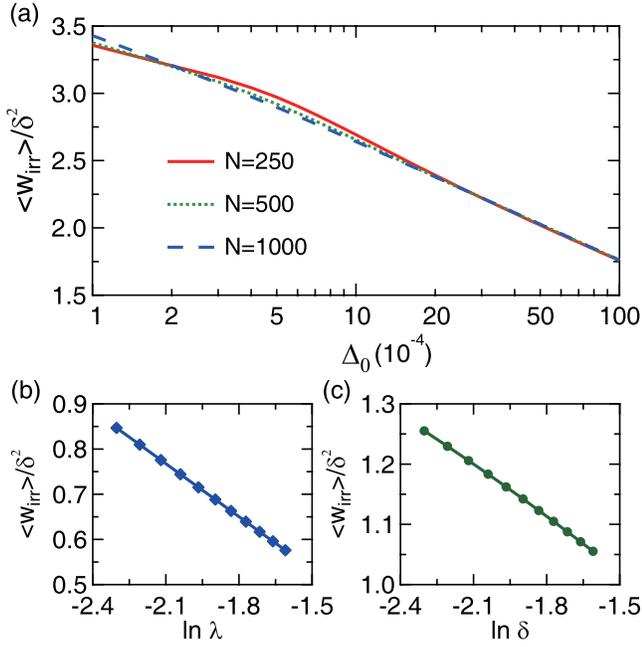}
\end{center}
\caption{(Color online) (a) The mean irreversible work density, $\langle w_{\text{irr}}\rangle /\delta^{2}$, as a function of the initial order parameter $\Delta_0$ of the BCS model in a single quench protocol for different values of the lattice size $N$. (b) and (c) plot its logarithmic scaling with respect to $\delta$ and $\lambda$ in the thermodynamic limit with $N=1000$. Other parameters are $\delta=0.01$ in (b) and $\lambda=0.01$ in (c).}
\label{fig8BCS-scaling}
\end{figure}

Now we extend our approach to the superconductivity. Since the model is generally insoluble, we consider here the mean-field BCS Hamiltonian with a bilinear form and discuss later in section \ref{con} the possibility of going beyond the mean-field approximation. The Hamiltonian is \cite{Altland2010},
\begin{equation}
H=\sum_{k}\Psi_{k}^{\dagger}\left[
\begin{array}
[c]{cc}
\xi_{k} & -\Delta\\
-\Delta & -\xi_{k}
\end{array}
\right]  \Psi_{k}, \label{BCS Hamiltonian}
\end{equation}
where $\xi_{k}=-2v [  \cos(  k_{x})  +\cos(  k_{y})]  -4v^\prime\cos(  k_{x})  \cos(  k_{y})  -\mu$, with the chemical potential $\mu$, the mean-field order parameter $\Delta$ and the Nambu spinor $\Psi_{k}^{\dagger}=(  c_{k\uparrow}^{\dagger},c_{-k\downarrow})  $. The parameters $v$ and $v^\prime$ denote the nearest-neighbor and next-nearest-neighbor hoppings on a two-dimensional lattice. We have $k_{x}=2\pi m/N$ and $k_{y}=2\pi l/N $ with $m, l=1-N/2,\ldots,\,N/2$, where $N$ is the lattice size along both $x$ and $y$ directions. Hereafter we set $v=0.435$, $v^\prime=0.05$ and $\mu=0.5$.

\begin{figure}[ptb]
\begin{center}
\includegraphics[width=8.6cm]{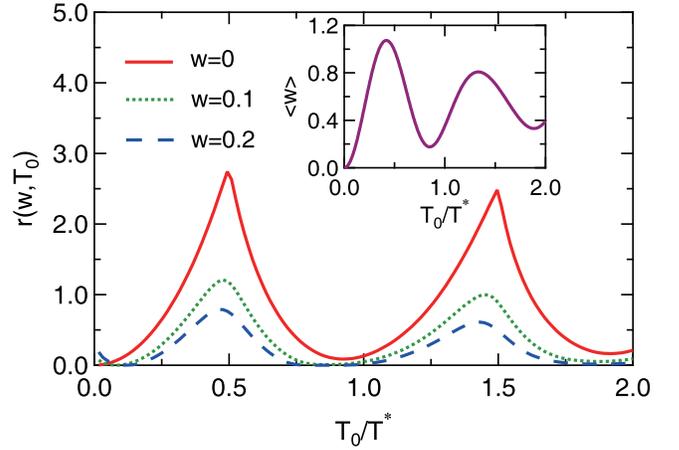}
\end{center}
\caption{(Color online) The rate function $r(  w,T_{0})  $ at zero temperature for a double quench of the BCS Hamiltonian from $\Delta_{0}=0.0$ to $\Delta_{1}=2.0$ and back to $\Delta_{0}$ after time $T_{0}$ for $N=50$. In contrast to the singular behavior at $T_0=(  n+1/2)  T^{\ast}$ in the rate function, the mean work density $\langle w\rangle $ shown in the inset exhibits no singularity from dynamical QPTs.}
\label{fig9BCSDQPT}
\end{figure}

For single quench, we tune the mean-field parameter $\Delta$ from an initial value $\Delta_{0}$ to $\Delta_{1}=\Delta_{0}+\delta$ such that $\mathbf{d}_{k}^{0}=(  -\Delta_{0},0,\xi_{k})  $ and $\mathbf{d}_{k}^{1}=(-\Delta_{1},0,\xi_{k})$. The irreversible work can then be calculated analytically (Appendix \ref{appC}). Figure \ref{fig8BCS-scaling}(a) plots $\langle w_{\text{irr}}\rangle /  \delta^{2}  $ as a function of $\Delta_{0}$ for different lattice size $N$. For large $N$, a logarithmic divergence appears as $\Delta_0$ approaches zero. In this thermodynamical limit ($N=\infty$), as shown in Figs. \ref{fig8BCS-scaling}(b) and \ref{fig8BCS-scaling}(c), we also obtain logarithmic scaling  with respect to $\lambda$ and $\delta$. This indicates that the critical exponents are $\nu=1/2$ and $z=2$ with $d=2$.

For the double quench protocol, we change the order parameter from $\Delta_{0}$ to $\Delta_{1}$ at $t=0$ and back to $\Delta_{0}$ at  $t=T_{0}$. Using the G\"{a}rter-Ellis theorem, we obtain the rate function $r(  w,T_{0})  $ in Fig. \ref{fig9BCSDQPT} with the initial order parameter $\Delta_{0}=0$. We see a clear nonanalytic behavior in $r(  w=0,T_{0})  $ at the critical times $t_c=(  n+1/2)T^{\ast}$, where $T^{\ast}=\pi/\Delta_1$. Once again, this indicates the existence of a dynamical QPT under double quench. Such a transition only occurs for $\Delta_0=0$ but is absent for any finite $\Delta_0$. It must be associated with the superconducting instability. We therefore speculate that the external driven field induces Cooper pair excitations around the initially free electron Fermi surface at $\Delta_0=0$. While the calculation of nonequilibrium dynamics might be otherwise involving for a general correlated Hamiltonian, extension of our approach to other strongly correlated phenomena is straightforward under the mean-field approximation.

\section{Discussion and Conclusions}

\label{con}

Our proposal of the functional field integral approach provides an alternate way to calculate the quantum work in an arbitrary time evolution protocol with a general bilinear Hamiltonian. The characteristic function and its cumulants contain all the major information for evaluating the work distribution via the G\"{a}rter-Ellis theorem and the mean (irreversible) work density, as well as the fidelity and Loschmidt echo in the double quench process. The applications of our approach to the transverse field Ising model, the SSH model, and the BCS model provide unambiguous evidences for signatures of quantum phase transitions and quantum criticality in the nonequilbirium process. In the sudden quench protocol, this is reflected in the quantum critical scaling of the mean irreversible work density, while in the double quench protocol, dynamical quantum phase transitions may be encoded as a singularity in the rate function. In the SSH model, we also see anomalous $1/N$ corrections due to the topological nature of the quantum phase transition. Compared to the Hamiltonian approach, the functional field integral formalism has the advantage to avoid tedious calculations of the second-order time differential equation and replace them by matrix products at different time intervals, which may be computed efficiently using optimized algorithms at the cost of explicit time evolution of the many-body wave function. As an example, we calculate the rate function for the SSH model with different time evolution protocols and find dynamical quantum phase transitions in all cases. However, the time periodicity only appears in the double quench case. Our approach provides a possibility for revealing the true driving parameter of the dynamical transition under general nonequilibrium conditions.

The transverse Ising model represents the few examples that can be mapped to a simple bilinear form. Often, an interacting many-body system cannot be solved even under equilibrium conditions, and approximations or numerical simulations have to be used with specially designed algorithms. These include the Hubbard model for cuprates and pnictides and the periodic Anderson model or the Kondo lattice model for heavy fermions. In these cases, certain auxiliary fields, based also on the functional field integral formalism, might be introduced to decouple the interacting terms in the Hamiltonian, so that the model can be solved under the mean-field (or saddle-point) approximations or simulated exactly using efficient numerical techniques \cite{Gubernatis2016}. The mean-field approximation can often capture some of the essential physics of the correlated model. As shown for the BCS Hamiltonian, a dynamical quantum phase transition is observed in a double quench process, which might be an intrinsic property of the model ascribed to the formation of Copper pairs near the Fermi surface driven by the time dependent pairing force.

To go beyond the mean-field approximation, one may consider Monte Carlo simulations of the auxiliary fields. A simpler situation is that the electrons are coupled to classical fields or subject to disorders but the Hamiltonian remains a bilinear form for each field or disorder configuration, as is in the Anderson localization model or the Ising Kondo lattice model. For the latter, we may have $H(t)=-v\sum_{\langle ij\rangle,\sigma}(  c_{i\sigma}^{\dagger}c_{j\sigma}+\text{h.c.})+J(t)\sum_{j\sigma}S_{j}^{z}\sigma c_{j\sigma}^{\dagger}c_{j\sigma}$, where $S_j^z$ is the Ising spin defined at site $j$ and $J(t)$ is the time-dependent coupling. Since all $S_j^z$ commute with $H$, the spins do not evolve with time. For each spin configuration, the electron Hamiltonian has a bilinear form in real space, $H(\{S_j^z\},t)=\sum_{i\sigma,l\sigma^\prime}c_{i\sigma}^{\dagger}A_{i\sigma,l\sigma^\prime}(\{S_j^z\},t)c_{l\sigma^\prime}$. Thus our derivation in appendix \ref{appA} can be extended to give a similar formula after the electron degrees of freedom are integrated out,
\begin{equation}
G(u)=\frac{\sum_{\{S_j^z\}}\det\left[  I+B(\{S_j^z\},T_{0})\right]}{\sum_{\{S_j^z\}}\det\left[  I+B(\{S_j^z\},0)\right]},
\end{equation}
where $B$ is defined as in Eq.~(\ref{Bk}) but with $A_k(t)$ replaced by the matrix $A(\{S_j^z\},t)$ for each spin configuration $\{S_j^z\}$. Then the Monte Carlo simulations can be applied on the Ising spin configurations \cite{Zhong2019}, with matrix products in real space instead of momentum space due to the lack of translation symmetry. Similar procedure may be extended to more complicated cases such as the Hubbard model or periodic Anderson model, where the local Coulomb interaction may be decoupled by introducing an auxiliary field $s_i$ at each site and time slice such that $e^{-\Delta\tau U(n_{i\uparrow}-1/2)(n_{i\downarrow}-1/2)}=c\sum_{s_{i}=\pm1}e^{\alpha s_{i}(n_{i\uparrow}-n_{i\downarrow})}$, where $c$ and $\alpha$ are constants depending on $\Delta\tau U$. Then one may perform Monte Carlo sampling on the auxiliary field configurations \cite{Gubernatis2016}. In both cases, detailed numerical realization may be limited by severe sign problem or other issues and thus require careful design of the algorithms. Nevertheless, our approach seems to provide a general scheme for the exploration of quantum work in nonequilibrium correlated systems. One interesting topic along this line would be the study of many-body localization within the framework of work statistics.

\section*{ACKNOWLEDGEMENTS}
We thank H. T. Quan for discussions. This work was supported by the National Natural Science Foundation of China (NSFC Grant Nos. 11774401, 11522435), the National Key R\&D Program of China (Grant No. 2017YFA0303103), the State Key Development Program for Basic Research of China (Grant No. 2015CB921303), and the National Youth Top-Notch Talent Support Program of China.

\onecolumngrid
\appendix

\section{Functional field integral approach to the characteristic function}
\label{appA}

The Hamiltonian we consider here is
\begin{equation}
H\left(  t\right)  =\sum_{k}\Psi_{k}^{\dagger}A_{k}\left(  t\right)  \Psi_{k},
\end{equation}
where $A_{k}(  t)  =\mathbf{d}_{k}(  t)  \cdot
\boldsymbol{\sigma}$ and $\boldsymbol{\sigma}$ is the vector of the Pauli
matrices. Due to the Dirac delta in the work distribution function in Eq. (\ref{wdf}), instead of calculating $p(w)$, it is often easier first to calculate its Fourier transformation, namely the characteristic function, $G(u)  =\int_{-\infty}^{\infty}dw\,\text{e}^{ \operatorname*{i}uwN^d}  p(w)$. Because of translational invariance, we have $G(u)=\prod\limits_{k}G_{k}(u)$ with
\begin{equation}
G_{k}(u)  =\frac{\operatorname*{Tr}\left[  U_{k}\left(
0,T_{0}\right)  \operatorname{e}^{\operatorname*{i}uH_{k}\left(  T_{0}\right)
}U_{k}\left(  T_{0},0\right)  \operatorname{e}^{-\left(  \operatorname*{i}%
u+\beta\right)  H_{k}\left(  0\right)  }\right]  }{\operatorname*{Tr}\left[
\operatorname{e}^{-\beta H_{k}\left(  0\right)  }\right]  }, \label{Gku}%
\end{equation}
and
\begin{equation}
U_{k}\left(  T_{0},0\right)  =\lim_{M\rightarrow\infty}\operatorname{e}%
^{-\operatorname*{i}H_{k}\left(  t_{M,M+1}\right)  \Delta t}\cdots
\operatorname{e}^{-\operatorname*{i}H_{k}\left(  t_{1,2}\right)  \Delta t},
\end{equation}
where $T_0$ is divided into $M$ slices and $\Delta t=T_{0}/M$ is an infinitesimal time step as illustrated in Fig.~\ref{fig1}. One may then write down the characteristic function by inserting a series of overcomplete bases of the fermionic coherent states on a closed time contour ($0\rightarrow T_{0}\rightarrow0$)\cite{Kamenev2011},
\begin{equation}
\widehat{1}=\int d\left[  \overline{\psi}^{j},\psi^{j}\right]
\operatorname{e}^{-\overline{\psi}^{j}\psi^{j}}\left\vert \psi^{j}%
\right\rangle \left\langle \psi^{j}\right\vert ,
\end{equation}
where $\vert\psi^j\rangle$ is defined at $t_j$ with $j=1,2,\ldots,2M+2$ on the contour. This yields
\begin{align}
{G}_{k}\left(  u\right)  &  =\frac{1}{\operatorname*{Tr}\left[  \operatorname{e}^{-\beta H_{k}\left(  0\right)}\right]  }
\int\left(  {\displaystyle\prod\limits_{n=1}^{2M+2}}d\left[\overline{\psi}_{k}^{n},\psi_{k}^{n}\right]  \right)  \exp\left(  -\sum_{j=1}^{2M+2}\overline{\psi}_{k}^{j}\psi_{k}^{j}\right)  \nonumber\\
&  \quad\times\left\langle -\psi_{k}^{2M+2}\left\vert \operatorname{e}^{\operatorname*{i}H_{k}\left(  t_{2M+1,2M+2}\right)  \Delta t}\right\vert
\psi_{k}^{2M+1}\right\rangle \cdots\left\langle \psi_{k}^{M+3}\left\vert\operatorname{e}^{\operatorname*{i}H_{k}\left(  t_{M+2,M+3}\right)  \Delta t}\right\vert \psi_{k}^{M+2}\right\rangle \left\langle \psi_{k}^{M+2}\left\vert \operatorname{e}^{\operatorname*{i}uH_{k}\left(  T_{0}\right)
}\right\vert \psi_{k}^{M+1}\right\rangle \nonumber\\
&  \quad\times\left\langle \psi_{k}^{M+1}\left\vert \operatorname{e}^{-\operatorname*{i}H_{k}\left(  t_{M,M+1}\right)  \Delta t}\right\vert
\psi_{k}^{M}\right\rangle \cdots\left\langle \psi_{k}^{2}\left\vert\operatorname{e}^{-\operatorname*{i}H_{k}\left(  t_{1,2}\right)  \Delta
t}\right\vert \psi_{k}^{1}\right\rangle \left\langle \psi_{k}^{1}\left\vert\operatorname{e}^{-\left(  \operatorname*{i}u+\beta\right)  H_{k}\left(
0\right)  }\right\vert \psi_{k}^{2M+2}\right\rangle .\label{GknuA}
\end{align}
Using the definition $\langle \psi_{k}^{m}\vert \psi_{k}^{n}\rangle=\exp\left(  \overline{\psi}_{k}^{m}\psi_{k}^{n}\right)$, we have for any $t$,
\begin{equation}
\left\langle \psi_{k}^{m}\left\vert \operatorname{e}^{\alpha H_{k}\left(  t\right)  \Delta t}\right\vert \psi_{k}^{n}\right\rangle
\approx\exp\left(  \overline{\psi}_{k}^{m}\operatorname{e}^{\alpha \Delta tA_{k}\left(  t\right)  }\psi_{k}^{n}\right)  .
\end{equation}
Thus the partition function at a single model $k$ may be reformulated as
\begin{align}
Z_{k}(0) &=\operatorname*{Tr}\left[  \operatorname{e}^{-\beta H_{k}\left(  0\right)  }\right] =\int d\left[  \overline{\psi},\psi\right]\operatorname{e}^{-\overline{\psi}\psi}\left\langle -\psi\left\vert\operatorname{e}^{-\beta H_{k}\left(  0\right)  }\right\vert \psi\right\rangle\nonumber\\
&=\int d\left[  \overline{\psi},\psi\right]  \exp\left[  -\overline{\psi}\left(  I+\operatorname{e}^{-\beta A_{k}\left(  0\right)  }\right)\psi\right]  =\det\left(  I+\operatorname{e}^{-\beta A_{k}\left(  0\right)}\right) \nonumber\\
&=2+\operatorname*{Tr}\left[  \operatorname{e}^{-\beta A_{k}\left(  0\right)  }\right],
\label{Eq:Z}
\end{align}
and similarly,
\begin{equation}
G_{k}\left(  u\right)  =\frac{\det\left[  I+\operatorname{e}^{\operatorname*{i}\Delta tA_{k}\left(  t_{1,2}\right)  }\cdots\operatorname{e}^{\operatorname*{i}\Delta tA_{k}\left(  t_{M,M+1}\right)}\operatorname{e}^{\operatorname*{i}uA_{k}\left(  T_{0}\right)  }\operatorname{e}^{-\operatorname*{i}\Delta tA_{k}\left(  t_{M,M+1}\right)}\cdots\operatorname{e}^{-\operatorname*{i}\Delta tA_{k}\left(  t_{1,2}\right)  }\operatorname{e}^{-\left(  \operatorname*{i}u+\beta\right)A_{k}\left(  0\right)  }\right]  }{\det\left[  I+\operatorname{e}^{-\beta A_{k}\left(  0\right)  }\right]  },
\end{equation}
where $I$ is the 2$\times$2 identity matrix. In deriving above equations, we have used the integral
\begin{equation}
\int d\left[  \overline{\psi},\psi\right]  \exp\left(  -\overline{\psi}A\psi+\overline{\chi}\psi+\overline{\psi}\chi\right)  =\det\left(  A\right)  \exp\left(  \overline{\chi}A^{-1}\chi\right).
\end{equation}
The above formula can be further simplified by defining $B_{k}(T_{0})=C_{k}^{\dagger}(  T_{0})  \operatorname{e}^{\operatorname*{i}uA_{k}(  T_{0})  }C_{k}(  T_{0})\operatorname{e}^{-(  \operatorname*{i}u+\beta)  A_{k}(0)  }$, with $C_{k}(  T_{0})  =\mathcal{T}\exp\left[  -\operatorname*{i}\int_{0}^{T_{0}}dtA_{k}(  t)\right]$ for arbitrary time dependence of $A_k(t)$ between 0 and $T_0$. Note that $B_k(0)=\operatorname{e}^{-\beta  A_{k}(0)}$. We have eventually
\begin{equation}
G(  u)  ={\displaystyle\prod\limits_{k}}\frac{\det\left[  I+B_{k}(T_{0})\right]  }{\det\left[  I+B_k(0)\right]  }={\displaystyle\prod\limits_{k}}\frac{2+\operatorname*{Tr}\left[  B_{k}\left(  T_{0}\right)  \right]
}{2+\operatorname*{Tr}\left[  \operatorname{e}^{-\beta A_{k}\left(  0\right)}\right]  },
\label{Guappendix}
\end{equation}
following the identity, $\det(  I+A)  =1+\operatorname*{Tr}A+\det A$, for any 2$\times$2 matrix $A$. The mean work density is given by the first cumulant of the characteristic function, $\langle w\rangle =-\operatorname*{i}dG(u)  /(N^d du)|_{u=0}$, yielding
\begin{equation}
\left\langle w\right\rangle =\frac{1}{N^d}\sum_{k}\frac{\operatorname*{Tr}\left[  \left(
D_{k}\left(  T_{0}\right)  -A_{k}\left(  0\right)  \right)  \operatorname{e}%
^{-\beta A_{k}\left(  0\right)  }\right]  }{2+\operatorname*{Tr}\left[
\operatorname{e}^{-\beta A_{k}\left(  0\right)  }\right]  },
\label{workgeneral}%
\end{equation}
where $D_{k}(  T_{0})  =C_{k}^{\dagger}(  T_{0})A_{k}(  T_{0})  C_{k}(  T_{0})  $. On the other hand, taking $u=\operatorname*{i}\beta$, one immediately derives the Jarzynski equality
\cite{Jarzynski1997PRL,Jarzynski1997PRE},
\begin{equation}
\left\langle \operatorname{e}^{-\beta w N^d}\right\rangle=\left\langle \operatorname{e}^{-\beta W}\right\rangle =
{\displaystyle\prod\limits_{k}}
\frac{2+\operatorname*{Tr}\left[  \operatorname{e}^{-\beta A_{k}\left(
T_{0}\right)  }\right]  }{2+\operatorname*{Tr}\left[  \operatorname{e}^{-\beta
A_{k}\left(  0\right)  }\right]  }=\frac{Z\left(  T_{0}\right)  }{Z\left(  0\right)}=\operatorname{e}^{-\beta\Delta F},
\end{equation}
where $\Delta F=F_{1}-F_{0}$ is the free energy difference between the final and initial equilibrium states at the inverse temperature
$\beta$, $Z(  0)  $ and $Z(  T_{0})  $ are the partition functions at $t=0$ and $t=T_{0}$, respectively. Recently, the quantum Jarzynski equality has been experimental verified in trapped ion system \cite{An2015,Xiong2018}.

\section{The work statistics and dynamical quantum phase transition}
\label{appB}

For a single quench where the model Hamiltonian changes from $A_{k}^{0}=\mathbf{d}_{k}^{0}\cdot\boldsymbol{\sigma}$ at $t=0^{-}$ to $A_{k}^{1}=\mathbf{d}_{k}^{1}\cdot\boldsymbol{\sigma}$ at $t=T_{0}=0^{+}$, the characteristic function is
\begin{equation}
G(  u)  ={\displaystyle\prod\limits_{k}}
\frac{2+\operatorname*{Tr}\left[  \operatorname{e}^{\operatorname*{i}%
uA_{k}^{1}}\operatorname{e}^{-\left(  \operatorname*{i}u+\beta\right)
A_{k}^{0}}\right]  }{2+\operatorname*{Tr}\left[  \operatorname{e}^{-\beta
A_{k}^{0}}\right]  }. \label{Chara}%
\end{equation}
An arbitrary matrix $A_{k}=\mathbf{d}_{k}\cdot\boldsymbol{\sigma}$ with $\mathbf{d}_{k}=(  x_{k},y_{k},z_{k})  $ may be diagonalized under the unitary transformation, $U_{k}^{-1}A_{k}U_{k}=D_{k}$, where
\begin{equation}
U_{k}=\left[
\begin{array}
[c]{cc}%
\mu_{k} & -\nu_{k}^{\ast}\\
\nu_{k} & \mu_{k}%
\end{array}
\right]  ,\quad D_{k}=\left[
\begin{array}
[c]{cc}%
E_{k} & 0\\
0 & -E_{k}%
\end{array}
\right]  ,
\end{equation}
with $E_{k}=\vert \mathbf{d}_{k}\vert$, $\mu_{k}=\sqrt{(E_{k}+z_{k})/2E_{k}}$, and $\nu_{k}=\frac{x_{k}+\operatorname*{i}y_{k}}
{\sqrt{x_{k}^{2}+y_{k}^{2}}}\sqrt{\frac{E_{k}-z_{k}}{2E_{k}}}$. The matrix exponential $\operatorname{e}^{\alpha A_{k}}$ is then
\begin{equation}
\operatorname{e}^{\alpha A_{k}}=U_{k}\operatorname{e}^{\alpha D_{k}}%
U_{k}^{-1}=\left[
\begin{array}
[c]{cc}%
\frac{E_{k}\cosh\left(  \alpha E_{k}\right)  +z_{k}\sinh\left(  \alpha
E_{k}\right)  }{E_{k}} & \frac{\left(  x_{k}-\operatorname*{i}y_{k}\right)
\sinh\left(  \alpha E_{k}\right)  }{E_{k}}\\
\frac{\left(  x_{k}+\operatorname*{i}y_{k}\right)  \sinh\left(  \alpha
E_{k}\right)  }{E_{k}} & \frac{E_{k}\cosh\left(  \alpha E_{k}\right)
-z_{k}\sinh\left(  \alpha E_{k}\right)  }{E_{k}}%
\end{array}
\right].
\end{equation}
We have
\begin{equation}
G_{k}\left(  u\right)  =\frac{1}{2}\operatorname{sech}^{2}\left(  \frac{\beta
E_{k}^{0}}{2}\right)  \left\{  1+\cos\left(  uE_{k}^{1}\right)  \cos\left[
E_{k}^{0}\left(  u-\operatorname*{i}\beta\right)  \right]  +\frac
{\mathbf{d}_{k}^{0}\cdot\mathbf{d}_{k}^{1}}{E_{k}^{0}E_{k}^{1}}\sin\left(
uE_{k}^{1}\right)  \sin\left[  E_{k}^{0}\left(  u-\operatorname*{i}%
\beta\right)  \right]  \right\}  . \label{charageneral}%
\end{equation}
where $E_{k}^{0,1}=\vert \mathbf{d}_{k}^{0,1}\vert $.

From Eq.~(\ref{Eq:Z}), the free energy density difference between the final and initial states in a quench process can be written as,
\begin{equation}
\Delta f= \frac{\Delta F}{N^d}=-\frac{1}{\beta N^d}\sum_{k} \ln \frac{2+\operatorname*{Tr}\left[
\operatorname{e}^{-\beta A_{k}\left(  T_{0}\right)  }\right]  }%
{2+\operatorname*{Tr}\left[  \operatorname{e}^{-\beta A_{k}\left(  0\right)
}\right]  }=-\frac{1}{\beta N^d}\sum_{k}\ln\frac{\cosh^{2}\left(  \beta E_{k}%
^{1}/2\right)  }{\cosh^{2}\left(  \beta E_{k}^{0}/2\right)  }.
\end{equation}
From Eq.~(\ref{workgeneral}) and considering $D_k(T_0)=A_k^1$ for the quench protocol, we have the mean work density,
\begin{equation}
\left\langle w\right\rangle =\frac{1}{N^d}\sum_{k}\frac{\operatorname*{Tr}\left[  \left(  A_{k}^{1}%
-A_{k}^{0}\right)  \operatorname{e}^{-\beta A_{k}^{0}}\right]  }%
{2+\operatorname*{Tr}\left[  \operatorname{e}^{-\beta A_{k}^{0}}\right]  }
=\frac{1}{N^d}\sum_{k}\frac{\left[  \left(  E_{k}^{0}\right)
^{2}-\mathbf{d}_{k}^{0}\cdot\mathbf{d}_{k}^{1}\right]  \tanh\left(  \frac
{1}{2}\beta E_{k}^{0}\right)  }{E_{k}^{0}}.
\end{equation}
Combing above equations gives the mean irreversible work density, $\langle w_{\text{irr}}\rangle =\langle w\rangle -\Delta f$. For a small quench, $\mathbf{d}_{k}^{0}=(  x_{k}^{0},y_{k}^{0}, z_{k}^{0})  $ $\rightarrow$ $\mathbf{d}_{k}^{1}=(  x_{k}^{0}+\delta,y_{k}^{0},z_{k}^{0})  $, we have the expansion,
\begin{equation}
E_{k}^{1}=E_{k}^{0}+\frac{x_{k}^{0}\delta}{E_{k}^{0}}+\frac{\left[  \left(
y_{k}^{0}\right)  ^{2}+\left(  z_{k}^{0}\right)  ^{2}\right]  \delta^{2}%
}{2\left(  E_{k}^{0}\right)  ^{3/2}}+O\left(  \delta^{3}\right).
\end{equation}
Thus at zero temperature
\begin{equation}
\left\langle w\right\rangle =\frac{1}{N^d}\sum_{k}\frac{-x_{k}^{0}\delta}{E_{k}^{0}}%
=-\frac{1}{N^d}\sum_{k}\delta\left.\frac{\partial E_{k}}{\partial x_{k}}\right\vert_{x_{k}=x_{k}^{0}},
\end{equation}%
\begin{equation}
\left\langle w_{\text{irr}}\right\rangle =\frac{1}{N^d}\sum_{k}\left(  \frac{-x_{k}%
^{0}\delta}{E_{k}^{0}}+E_{k}^{1}-E_{k}^{0}\right)  =\frac{1}{N^d}\sum_{k}\frac{\delta^{2}%
}{2}\left.\frac{\partial^{2}E_{k}}{\partial x_{k}^{2}}\right\vert_{x_{k}=x_{k}^{0}}.
\end{equation}
We see that the singularity in $\left\langle w\right\rangle $ and $\left\langle w_{\text{irr}}\right\rangle $ reflects the first and second-order phase transitions, respectively.

The work distribution $p(w)$ may be calculated using the Fourier transformation of $G(u)$. In general, the energy density difference of the ground states, $\Delta_{\min}=\Delta f=-\frac{1}{N^d}\sum_{k}\left[  E_{k}^{1}-E_{k}^{0}\right]  $ gives the minimal work that can be measured at zero temperature, i.e., $p(w<\Delta_{\min})=0$ for $T=0$. While at finite temperatures, even the highest excited state may have a small but finite weight due to thermal effect. Thus the minimal work $w_{\min}$ that can be achieved is given by the energy density difference between the ground state of the postquench Hamiltonian with energy density $\epsilon^{g}=-\frac{1}{N^{d}}\sum_{k}E_{k}^{1}$ and the highest excited state of the initial Hamiltonian with the energy density $\epsilon^{h}=\frac{1}{N^{d}}\sum_{k}E_{k}^{0}$, yielding $w_{\min}=-\frac{1}{N^{d}}\sum_{k}(E_{k}^{1}+E_{k}^{0})$. Oppositely, the maximal work $w_{\max}$ is given by the energy density difference between the highest excited state of the postquench Hamiltonian with energy density $\epsilon^{h}=\frac{1}{N^{d}}\sum_{k}E_{k}^{1}$ and the ground excited state of the initial Hamiltonian with the energy density $\epsilon^{g}=-\frac{1}{N^{d}}\sum_{k}E_{k}^{0}$, yielding $w_{\max}=\frac{1}{N^{d}}\sum_{k}(E_{k}^{1}+E_{k}^{0})=-w_{\min}$. Beyond the interval $[w_{\min}, w_{\max}]$, the distribution function $p(w)  =0$ and the rate function $r(w) =\infty$ at all temperatures.

For a double quench, where the Hamiltonian changes from $A_{k}^{0}$ to $A_{k}^{1}$ at $t=0$ and back to $A_{k}^{0}$ for $t\geq T_{0}$, we have $C_{k}(T_{0})=\operatorname{e}^{-\operatorname*{i} T_{0} A_{k}^{1}}$. The characteristic function at zero temperature can be evaluated to be
\begin{equation}
G(u)  =\prod_{k}\operatorname{e}^{\operatorname*{i}uE_{k}^{0}}\left\{
\cosh(-\operatorname*{i}uE_{k}^{0})+\sinh(-\operatorname*{i}uE_{k}^{0})\left[
\cos^{2}(E_{k}^{1}T_{0})+\frac{2(\mathbf{d}_{k}^{0}\cdot\mathbf{d}_{k}%
^{1})^{2}-\left(  E_{k}^{1}E_{k}^{0}\right)  ^{2}}{\left(  E_{k}^{1}E_{k}%
^{0}\right)  ^{2}}\sin^{2}(E_{k}^{1}T_{0})\right]  \right\}  .
\end{equation}
Because $c(R)=-\lim_{N\rightarrow\infty}\frac{1}{N^d}\ln G(u=\operatorname*{i}R)$ and $r(w=0)=\sup_{R\in\mathbb{R}}c( R)$, the singularity in the rate function is associated with the roots of $G(R)$. Thus a dynamical quantum phase transition may occur at the critical time, $t_{c}=\frac{\left(2n+1\right)  \pi}{2E_{k_{c}}^{1}}$ ($n=0$, 1, 2, ...) if there exits a critical momentum $k_{c}$ satisfying $\mathbf{d}_{k_{c}}^{0}\cdot\mathbf{d}_{k_{c}}^{1}=0$. Under this condition, we have at $T_0=t_c$,
\begin{equation}
G_{k_{c}}(R)=\operatorname{e}^{-RE_{k_{c}}^{0}}\left\{  \cosh(RE_{k_{c}}%
^{0})+\sinh(RE_{k_{c}}^{0})\left[  \cos^{2}(n+\frac{\pi}{2})-\sin^{2}%
(n+\frac{\pi}{2})\right]  \right\}  =\operatorname{e}^{-2RE_{k_{c}}^{0}}.
\end{equation}
Then the singularity occurs when $G(R)  =\prod_{k}G_{k}(R)=0$ as $R\rightarrow\infty$. While for all other $T_0\neq t_c$ or $k\neq k_c$, there exists a term in the brace proportional to $\operatorname{e}^{RE_{k_{c}}^{0}}$, which will cancel the prefactor $\operatorname{e}^{-RE_{k}^{0}}$ and produce a nonzero $G(R)$ for all $R$.

\section{Application to the models}
\label{appC}

For the transverse Ising chain, we have $\mathbf{d}_{k}=(0,-2\sin k,-2h-2\cos k)$. Using the dispersion relation $\epsilon_{k}(  h)  =\vert \mathbf{d}_{k}\vert =2\sqrt{1+h^{2}+2h\cos k}$, we can calculate $\left\langle w_{\text{irr}}\right\rangle$ for the single quench from $h_0<1$ to $h_0>1$,
\begin{equation}
\left\langle w_{\text{irr}}\right\rangle =\frac{1}{N}\sum_{k>0}\left[  \frac
{-4\delta\left(  h_{0}+\cos k\right)  }{\epsilon_{k}\left(  h_{0}\right)
}+\epsilon_{k}\left(  h_{1}\right)  -\epsilon_{k}\left(  h_{0}\right)
\right]  .
\end{equation}
Using $t_{c}=\frac{(2n+1)  \pi}{2E_{k_{c}}^{1}}$ and $\mathbf{d}_{k_{c}}^{0}\cdot\mathbf{d}_{k_{c}}^{1}=0$ for the value of $k_c$, the critical times for the dynamical QPT during a double quench process are
\begin{equation}
t_{c}=\frac{\left(  n+1/2\right)  \pi}{2\sqrt{1+\left(  h_{1}\right)
^{2}-2h_{1}\frac{1+h_{0}h_{1}}{h_{0}+h_{1}}}}.
\end{equation}

For the SSH model, we have $\mathbf{d}_{k}=\left(  v+\cos k,\sin k,0\right)
$ and $\epsilon_{k}\left(  v\right)  =$ $\left\vert \mathbf{d}_{k}\right\vert
=\sqrt{1+v^{2}+2v\cos k}$. Thus for a single quench from $v_0<1$ to $v_1=v_0+\delta>1$,
\begin{equation}
\left\langle w_{\text{irr}}\right\rangle =\frac{1}{N}\sum_{k}\left[\frac{-\delta\left(
v_{0}+\cos k\right)  }{\epsilon_{k}\left(  v_{0}\right)  }+\epsilon_{k}\left(
v_{1}\right)  -\epsilon_{k}\left(  v_{0}\right) \right ].\label{Wirrssh}%
\end{equation}
Because of the topological nature of the QPT in the SSH model, there exists a band crossing at $k=\pi$ at the QCP. We can thus separate the mean irreversible work density into two terms, $\langle w_{\text{irr}}\rangle=\langle w_{\text{irr}}\rangle_{k \neq \pi}+\left\langle w_{\text{irr}}\right\rangle_{k=\pi}$. The results are shown in Fig.~\ref{figsshN2}. While the first term gives the usual $\ln N$ scaling due to quantum criticality, the second term $\langle w_{\text{irr}}\rangle_{k=\pi}=\frac{2}{N}\left(  \delta-\lambda\right)$ yields an additional $1/N$ contribution, which becomes dominant and leads to the anomalous $1/N$ scaling when $\lambda ^{-\nu}$ is the largest length scale. For the double quench process, the critical times for the dynamical QPT are
\begin{equation}
t_{c}=\frac{\left(  n+1/2\right)  \pi}{\sqrt{1+\left(  v_{1}\right)
^{2}-2v_{1}\frac{1+v_{0}v_{1}}{v_{0}+v_{1}}}}.
\end{equation}

For the BCS model with $\mathbf{d}_{k}=(-\Delta,0,\xi_{k})  $, the mean irreversible work density for the single quench process is
\begin{equation}
\left\langle w_{\text{irr}}\right\rangle =\frac{1}{N^2}\sum_{k} \left[ \frac{-\delta\Delta_{0}%
}{\sqrt{\Delta_{0}^{2}+\xi_{k}^{2}}}+\sqrt{\Delta_{1}^{2}+\xi_{k}^{2}}%
-\sqrt{\Delta_{0}^{2}+\xi_{k}^{2}} \right ],
\end{equation}
and the dynamical QPT during a double quench process with $\Delta_0=0$ occurs at
\begin{equation}
t_{c}=\frac{\left(  n+1/2\right)  \pi}{\Delta_{1}}.
\end{equation}

\twocolumngrid

\end{document}